\documentclass[11pt]{article}
\usepackage{amsmath,amssymb,amscd}

\usepackage{cite}
\usepackage{amsmath, amsthm, amssymb, amsfonts}
\usepackage{dsfont}
\usepackage{float}
\usepackage[a4paper]{geometry}
\usepackage{color}

\definecolor{red}{rgb}{1,0,0}

\usepackage{graphicx}
\usepackage[mathscr]{eucal}

\large\normalsize

\title{Noise-induced stochastic Nash equilibrium}
\author{Cong Li$^{1,\#}$, Tianjiao Feng$^{2,3,\#}$, Xiudeng Zheng$^{2, \#}$,
Sabin Lessard$^{4}$\footnote{Author for correspondence, and e-mail: lessards@dms.umontreal.ca} and Yi
Tao$^{1,2,5}$\footnote{Author for correspondence, and e-mail:
yitao@ioz.ac.cn}
\\
$^1$School of Ecology and Environment, Northwestern Polytechnical University,\\
Xi'an 710072, P.R. China \\
$^2$Key Laboratory of Animal Ecology and Conservation Biology,\\
Center for Computational and Evolutionary Biology, \\
Institute of Zoology, Chinese Academy of Sciences, \\
Beijing 100101, P.R. China \\
$^3$University of Chinese Academy of Sciences, \\ Beijing 100049, P.R. China \\
$^{4}$Department of Mathematics and Statistics, University of Montreal, \\
Montreal QC H3C 3J7, Canada \\
$^5$Institute of Biomedical Research, Yunnan University, \\
Kunming 650091, P.R. China \\
$^{\#}$These authors have the same contribution to this paper. }

\date{}

\begin{document}

\maketitle

\newpage

\section*{Abstract}

In order to better understand the impact of environmental stochastic
fluctuations on the evolution of animal behavior, we introduce the
concept of a stochastic Nash equilibrium (SNE) that extends the
classical concept of a Nash equilibrium (NE). Based on a stochastic
stability analysis of a linear evolutionary game with temporally
varying payoffs, we address the question of the existence of a SNE,
either weak when the geometric mean payoff against it is the same
for all other strategies or strong when it is strictly smaller for
all other strategies, and its relationship with a stochastically
evolutionarily stable (SES) strategy. While a strong SNE is always
SES, this is not necessarily the case for a weak SNE. We give
conditions for a completely mixed weak SNE not to be SES and to
coexist with at least two strong SNE. More importantly, we show that
a pair of two completely mixed strong SNE can emerge as the noise
level increases. This not only indicates that a noise-induced SNE
may possess some properties that a NE cannot possess, such as being
completely mixed and strong, but also illustrates the complexity of
evolutionary game dynamics in a stochastic environment.

\newpage
\emph{\textbf{Introduction.}} As it is well known, a Nash
equilibrium (NE) is the core concept of non-cooperative games
\cite{nash1950}, and it has had a profound impact on economics,
biology and social sciences
\cite{wei1997,smi1973,smi1974,smi1982,bro2022}. For linear
evolutionary games based on payoff matrices \cite{hof1998,bro2022},
the equilibrium condition for an evolutionarily stable strategy
(ESS) is exactly the definition of a NE, which is a strategy that is
the best reply to itself \cite{smi1982,hof1998}. It is also known in
this framework that, while an ESS must be a NE, the inverse is not
necessarily true. This is the case, however, for a strict NE, which
is strictly better against itself than any other strategy
\cite{smi1982,hof1998}. Moreover, if a completely mixed strategy is
a NE, then it must be unique and it can never be a strict NE. In
particular, this implies that it is impossible to have two or more
completely mixed strategies that are both ESS \cite{hof1998}.

Recently, in order to explore the impact of environmental stochastic
fluctuations on evolutionary game dynamics, Zheng \emph{et al.}
\cite{zhe2017,zhe2018} (see also Feng \emph{et al.}
\cite{fen2021,fen2022}) developed the concept of stochastic
evolutionary stability based on conditions for stochastic stability
of equilibria in stochastic recurrence equations (or stochastic
replicator dynamics). A stochastically evolutionarily stable (SES)
strategy is defined as a strategy such that, if all the members of
the population adopt it, then the probability for at least any
slightly perturbed strategy to successfully invade the population
under the influence of natural selection is arbitrarily low.

Then, a challenging question naturally arises: how should we define
a stochastic Nash equilibrium (SNE) in the case of random payoffs
that would extend the concept of a NE in the case of deterministic
payoffs, and what would be the relationships between a SNE and a SES
strategy in stochastic evolutionary games. Analogously to the
classic definition of a NE \cite{wei1997,hof1998,now2006,bro2022}, a
SNE should be defined as a strategy that is the best reply to itself
but taking into account the uncertainty in the payoffs in a
stochastic environment. Hence, a SNE should not only be regarded as
an extension of a NE, but also as a concept to capture the effect of
environmental noise on the equilibrium structure in evolutionary
game dynamics.

In this letter, from a stochastic stability analysis of the discrete-time dynamics of a linear evolutionary game with a random payoff matrix, we define the concepts of weak SNE and strong SNE, and we examine conditions for their existence and co-existence. Our goal is not only to show how stochastic environmental noise can induce the emergence of a SNE, called a noise-induced SNE, which does not have any equivalent in a constant environment, but also to provide a theoretical framework for studying the role of environment noise in shaping complex equilibrium structures and evolutionary patterns in game dynamics.

\emph{\textbf{Stochastic Nash equilibrium.}} We consider an
evolutionary game in an infinite population with discrete,
non-overlapping generations. There are two pure strategies in use,
denoted by $S_1$ and $S_2$, and the payoffs received following
pairwise interactions at time step $t \ge 1$ are given by the
entries of the game matrix
\begin{align}
\mathbf{A}(t) &= \begin{pmatrix} a_{11}(t) & a_{12}(t) \\ a_{21}(t) & a_{22}(t) \end{pmatrix} ,
\end{align}
where $a_{ij}(t)$ denotes the payoff to strategy $S_i$ against strategy $S_j$ at time step $t \ge 1$ for $i,j=1,2$. For simplicity,
these payoffs are assumed to be positive random variables that are uniformly bounded below and above by some positive constants. Therefore, there exist real numbers $A$ and $B$ such that $0<A \le a_{ij}(t) \le B$ for $i,j=1,2$ and all $t \ge 1$ \cite{zhe2017}. Moreover, the probability distribution of $a_{ij}(t)$ for $i,j=1,2$ do not depend on $t \ge 1$. The means, variances, and covariances of these random payoffs are given by $\left <a_{ij}(t) \right> = \bar{a}_{ij}$, $\left < \big (a_{ij}(t) -\bar{a}_{ij} \big)^2 \right> =\sigma_{ij}^2$ and $\left< \big (a_{ij}(t) -\bar{a}_{ij} \big) \big (a_{kl}(t) -\bar{a}_{kl} \big) \right> = \sigma_{ij,kl}$, respectively, for $i,j,k,l=1,2$ with $(i,j) \ne (k,l)$. As for $s \ne t$, the payoffs $a_{ij}(s)$ and $a_{kl}(t)$ are assumed to be
independent so that $\left< \big(a_{ij}(s) -\bar{a}_{ij} \big) \big ( a_{kl}(t) -\bar{a}_{kl} \big ) \right >=0$ for $i,j,k,l=1,2$.

Consider a population consisting of individuals using only two mixed strategies $\mathbf{x}=(x, 1-x)$ and $\hat{\mathbf{x}}=(\hat{x}, 1-\hat{x})$ with $x, \hat{x} \in [0,1]$. The payoff matrix for these two mixed strategies at time step $t \ge 1$ is given by
\begin{align}
\begin{pmatrix} \mathbf{x} \cdot \mathbf{A}(t) \mathbf{x} & \mathbf{x} \cdot \mathbf{A}(t) \hat{\mathbf{x}} \\ \hat{\mathbf{x}} \cdot \mathbf{A}(t) \mathbf{x} & \hat{\mathbf{x}} \cdot \mathbf{A}(t) \hat{\mathbf{x}} \end{pmatrix} ,
\end{align}
where $\mathbf{x} \cdot \mathbf{A}(t) \mathbf{x}$ (or $\mathbf{x} \cdot \mathbf{A}(t) \hat{\mathbf{x}}$) is the payoff to strategy
$\mathbf{x}$ against strategy $\mathbf{x}$ (or strategy $\hat{\mathbf{x}}$), and $\hat{\mathbf{x}} \cdot \mathbf{A}(t) \mathbf{x}$ (or $\hat{\mathbf{x}} \cdot \mathbf{A}(t) \hat{\mathbf{x}}$) is the payoff to strategy $\hat{\mathbf{x}}$ against strategy $\mathbf{x}$ (or strategy $\hat{\mathbf{x}}$) \cite{zhe2017}.

Let $q(t)$ be the frequency of strategy $\mathbf{x}$ at time step $t \ge 1$. Assuming random pairwise interactions, the average payoffs to strategies $\mathbf{x}$ and $\hat{\mathbf{x}}$ at time step $t \ge 1$ are given by $\pi_{\mathbf{x}}(t) = q(t) \mathbf{x} \cdot \mathbf{A}(t) \mathbf{x} +(1-q(t)) \mathbf{x} \cdot \mathbf{A}(t) \hat{\mathbf{x}}$ and $\pi_{\hat{\mathbf{x}}}(t) = q(t) \hat{\mathbf{x}} \cdot \mathbf{A}(t) \mathbf{x} +(1-q(t)) \hat{\mathbf{x}} \cdot \mathbf{A}(t) \hat{\mathbf{x}}$, respectively. Taking the average payoff as fitness, the frequency of strategy $\mathbf{x}$ at time step $t+1$ can be expressed as
\begin{align}
q(t+1) &= \frac{q(t) \pi_{\mathbf{x}}(t)}{q(t) \pi_{\mathbf{x}}(t)+(1-q(t)) \pi_{\hat{\mathbf{x}}}(t)} ,
\end{align}
which is a stochastic recurrence equation \cite{zhe2017}.

By definition, the strategy $\hat{\mathbf{x}}$ is \emph{stochastically evolutionarily stable} (SES) if the boundary equilibrium $q(t)=0$ is \emph{stochastically locally stable} (SLS) for all possible $\mathbf{x} \ne \hat{\mathbf{x}}$ \cite{zhe2017}. It can be shown that $\hat{\mathbf{x}}$ is SES if and only if
\begin{align}\label{SNE}
\left< \log \mathbf{x} \cdot \mathbf{A}(t) \hat{\mathbf{x}} \right >\le \left< \log \hat{\mathbf{x}} \cdot \mathbf{A}(t) \hat{\mathbf{x}} \right >
\end{align}
for all possible $\mathbf{x}$, and
\begin{align}\label{stability}
\left< \frac{\hat{\mathbf{x}} \cdot \mathbf{A}(t) \mathbf{x}}{\hat{\mathbf{x}} \cdot \mathbf{A}(t) \hat{\mathbf{x}}} \right > -\left< \frac{\mathbf{x} \cdot \mathbf{A}(t) \mathbf{x}}{\hat{\mathbf{x}} \cdot \mathbf{A}(t) \hat{\mathbf{x}}} \right >= -(\hat{x}-x)^2 D > 0
\end{align}
with
\begin{align}
D &= \left < \frac{\big( a_{11}(t) -a_{12}(t) -a_{21}(t) +a_{22}(t) \big )^2}{a_{11}(t)a_{22}(t)-a_{12}(t)a_{21}(t)} \right >
\end{align}
in the case of an equality in Eq. (\ref{SNE}) for all possible
$\mathbf{x}$ (see \cite{zhe2017} and the Appendix for the expression
of $D$).

By analogy with the conditions for equilibrium and stability of an
\emph{evolutionarily stable strategy} (ESS) (see p. 63 in
\cite{hof1998}), the first condition above is used to define a
\emph{stochastic Nash equilibrium} (SNE). This corresponds to a
strategy that is the best reply to itself in a stochastic
environment based on the geometric means of the payoffs rather than
their arithmetic means.

Let us recall that the geometric mean of a random variable $X$ is
defined as $GM\left<X\right>= \exp (\left< \log X \right>)$.
Therefore, Eq. (\ref{SNE}) is equivalent to
\begin{align}\label{SNEbis}
GM\left< \mathbf{x} \cdot \mathbf{A}(t) \hat{\mathbf{x}} \right> \le GM\left< \hat{\mathbf{x}} \cdot \mathbf{A}(t) \hat{\mathbf{x}} \right>
\end{align}
for all possible $\mathbf{x} \ne \hat{\mathbf{x}}$. This is the
condition for $\hat{\mathbf{x}} $ to be a SNE. In the special case
where $GM\left< \mathbf{x} \cdot \mathbf{A}(t) \hat{\mathbf{x}}
\right> = GM\left< \hat{\mathbf{x}} \cdot \mathbf{A}(t)
\hat{\mathbf{x}} \right>$ for all $\mathbf{x} \ne \hat{\mathbf{x}}$,
the strategy $\hat{\mathbf{x}} $ will be called a \emph{weak
stochastic Nash equilibrium} (weak SNE). At the other extreme, if
$GM\left< \mathbf{x} \cdot \mathbf{A}(t) \hat{\mathbf{x}} \right> <
GM\left< \hat{\mathbf{x}} \cdot \mathbf{A}(t) \hat{\mathbf{x}}
\right>$ for all $\mathbf{x} \ne \hat{\mathbf{x}}$, then
$\hat{\mathbf{x}} $ will be said a \emph{strong stochastic Nash
equilibrium} (strong SNE). Note that the SNE condition is necessary
but not sufficient for stochastic evolutionary stability (SES),
while the condition for a strong SNE is sufficient but not
necessary. In other words, we have the following implications:
$$\textrm{ strong SNE } \Rightarrow \textrm{ SES } \Rightarrow \textrm{ SNE }$$

\textbf{\emph{Equilibrium structure.}}
In this section, we examine the equilibrium structure of the system. 
In order to distinguish the pure strategies $S_1$ and $S_2$, we
assume throughout that $\mathbf{P} \left ( a_{11}(t)=a_{21}(t),
a_{22}(t)=a_{12}(t) \right)<1$, which is equivalent to saying that
$S_1$ and $S_2$ have different payoffs with positive probability. As
shown in the Appendix, a strategy $\mathbf{x^*}=(x^*, 1-x^*)$ such
that $\mathbf{P} \left ( \big( \mathbf{A}(t) \mathbf{x^*} \big )_1 =
\big( \mathbf{A}(t) \mathbf{x^*} \big )_2 \right)<1$ is a strong SNE
if and only if
\begin{align}\label{strongSNE}
\left< \frac{\big (\mathbf{A}(t) \mathbf{x^*} \big )_1 -\big ( \mathbf{A}(t)\mathbf{x^*} \big )_2}{\mathbf{x^*} \cdot \mathbf{A}(t) \mathbf{x^*}} \right >
\begin{cases}
\le 0& \textrm{ if } x^*=0,\\
\ge 0& \textrm{ if } x^*=1,\\
=0 &\textrm{ if } 0 < x^*<1,
\end{cases}
\end{align}
Moreover, if we have $\mathbf{P} \left ( \big( \mathbf{A}(t)
\mathbf{x} \big )_1 = \big( \mathbf{A}(t) \mathbf{x} \big )_2
\right)<1$ for all possible $\mathbf{x}$, then there exists at least
one strong SNE $\mathbf{x^*}=(x^*, 1-x^*)$, which is necessarily
SES.

On the other hand, if we have $\mathbf{P} \big ( \big (\mathbf{A}(t)
\hat{\mathbf{x}} \big )_1 = \big (\mathbf{A}(t) \hat{\mathbf{x}}
\big )_2 \big )=1$ for some $\hat{\mathbf{x}}=(\hat{x}, 1-\hat{x})$,
then $\hat{\mathbf{x}}$ is a weak SNE with an equality in Eq.
(\ref{SNE}) for all possible $\mathbf{x}$, and it is the unique weak
SNE in the system. In the case where $\mathbf{P} \left (
a_{11}(t)=a_{21}(t) \right)=1$, this unique weak SNE is
$\hat{\mathbf{x}}=(1, 0)$ and, owing to Eq. (\ref{stability}), it is
SES if
\begin{align}
D=\left <
\frac{a_{22}(t)-a_{12}(t)}{a_{11}(t)} \right> <0.
\end{align}
Analogously, in the case where $\mathbf{P} \left ( a_{22}(t)=a_{12}(t) \right)=1$, the unique weak SNE is $\hat{\mathbf{x}}=(0, 1)$, which is SES if
\begin{align}
D=\left <
\frac{a_{11}(t)-a_{21}(t) }{a_{22}(t)} \right> <0.
\end{align}
Finally, if there exists $r>0$ such that $\mathbf{P} \big (a_{11}(t)-a_{21}(t)=r\left(a_{22}(t)-a_{12}(t)\right) \big )=1$, then $\hat{\mathbf{x}}=(1/(1+r), r/(1+r))$ is the unique weak SNE. Moreover, if
\begin{align}
D = \left< \frac{a_{22}(t)-a_{12}(t)}{ ra_{22}(t)+a_{21}(t)} \right > <0,
\end{align}
then $\hat{\mathbf{x}}$ is SES.

In the case where $D>0$ in the above three cases, there exists at
least one strong SNE $\mathbf{x^*}=(x^*, 1-x^*)$ in the first two
cases, and even at least two strong SNE $\mathbf{x}_1^*=(x^*_1,
1-x^*_1)$ and $\mathbf{x}_2^*=(x^*_2, 1-x^*_2)$ in the third case.
As for $D=0$ in the above three cases, defining the quantity
\begin{align}
\hat{u}&= \left < \frac{\alpha(t)^3 \beta(t) }{\left( \det
\mathbf{A}(t) \right)^2} \right > \Bigg / \left <
\frac{\alpha(t)^4}{\left( \det \mathbf{A}(t) \right)^2} \right > \ ,
\end{align}
where $\alpha(t)=a_{11}(t)-a_{12}(t)-a_{21}(t)+a_{22}(t)$ and
$\beta(t)=a_{22}(t)-a_{21}(t)$, it can be shown that there is at
least one strong SNE $\mathbf{x}_1^*=(x^*_1, 1-x^*_1)$ with $x^*_1
\in [0, \hat{x})$ if $\hat{x} > \max (0, \hat{u})$, or $x^*_1 \in
(\hat{x}, 1]$ if $\hat{x} < \min (1, \hat{u})$.

\textbf{\emph{An example.}} In order to show how environmental noise can induce the emergence of a SNE, we now consider a specific example. Suppose a random payoff matrix at time step $t\ge 1$ in the form
\begin{align}
\mathbf{A}(t) &= \begin{pmatrix} \mu+a & \mu+a \\ \mu & \mu+\xi (t) \end{pmatrix} .
\end{align}
Here, $a$ and $\mu$ are positive constants with $\mu$ small enough
but $\mu \ne 0$, while $\xi (t)$ is a non-negative random variable
with $\xi (t)=b>a$ with probability $p$ and $\xi (t)=0$ with
probability $1-p$ ($0 <p<1$), so that $\bar{\xi}=\left<\xi
(t)\right>=pb$ and $\sigma_{\xi}^2=\left < (\xi(t) -\bar{\xi})^2
\right > =p(1-p)b^2$. Note that the mean payoff matrix
$\bar{\mathbf{A}}=\begin{pmatrix} \mu+a & \mu+a \\ \mu & \mu+pb
\end{pmatrix}$ corresponds to a stag-hunt game, or a coordination
game, if $pb>a$ \cite{bro2022}.

First, we find $\mathbf{P} \big ( \big (\mathbf{A}(t) \mathbf{x}
\big )_1 = \big (\mathbf{A}(t) \mathbf{x} \big )_2 \big )=\mathbf{P}
\big ((1-x) \xi(t)=a \big)<1$ and
\begin{align}
\left< \frac{\big (\mathbf{A}(t) \mathbf{x} \big )_1 -\big ( \mathbf{A}(t)\mathbf{x} \big )_2}{\mathbf{x} \cdot \mathbf{A}(t) \mathbf{x}} \right > =\left< \frac{a-(1-x)\xi(t)}{\mu+ax+(1-x)^2\xi(t)} \right >
\end{align}
for all possible $\mathbf{x}=(x, 1-x)$. Thus, owing to Eq. (\ref{strongSNE}), the strategy $\mathbf{x}^*=(1,0)$ is a strong SNE
since $a/(\mu+a)\ge 0$, while the strategy $\mathbf{x}^*=(0,1)$ is a strong SNE if and only if
\begin{align}
\left(\frac{a-b}{\mu+b}\right)p + \left(\frac{a}{\mu}\right)(1-p) \le 0 \ ,
\end{align}
which is equivalent to $p \ge \frac{a(\mu+b)}{b(\mu+a)}\in (0, 1)$.
As for a strong SNE $\mathbf{x}^*=(x^*, 1-x^*)$ with $x^* \in
(0,1)$, it must be the solution of the equation
\begin{align}
\left (\frac{a-(1-x^*)b}{\mu +ax^* +(1-x^*)^2 b} \right )p + \left( \frac{a}{\mu +ax^*}\right)(1-p) = 0 \ ,
\end{align}
which is the case if and only if
\begin{align}
abx^{*2} + \big [ a(a-b) +pb \mu -(1-p)ab \big ]x^* +\big [ p(a-b) \mu +(1-p) a (b+\mu) \big ] =0 \ .
\end{align}
Since $\mu$ is assumed to be small, the above equation can be approximated as
\begin{align}
bx^{*2} - \big ( (2-p)b-a \big) x^*+(1-p)b=0 \ ,
\end{align}
whose solutions are
\begin{align}
x_{1,2}^*=\frac{b(2-p)-a \pm \sqrt{(bp+a)^2-4ab}}{2b} \in (0,1)
\end{align}
under the condition that
\begin{align}
p\ge \frac{2 \sqrt{ab}-a}{b} \in (0,1) \ .
\end{align}
Therefore, for $p$ large enough, there may exist up to two strong
SNE besides $(0, 1)$ and $(1, 0)$ that do not exist for small $p$.
The results of stochastic simulations are shown in \textbf{Fig. 1},
and we can see that these results exactly match the theoretical
predictions.

\begin{figure}
    \begin{center}
    \includegraphics[height=7.2cm,width=8.5cm]{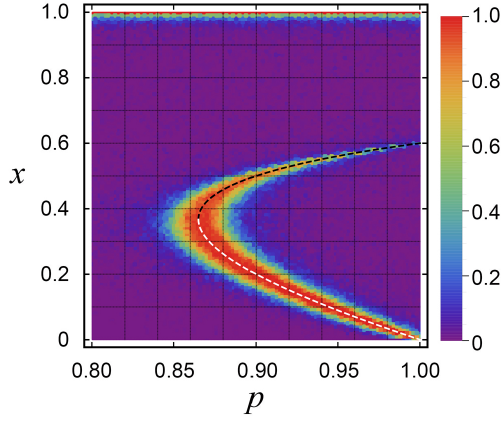}
    \end{center}
    \caption{\textbf{Stochastic simulation results for the existence of
a strong SNE in the example.} We take $b=10$, $c=4$ and $\mu=0.01$
in the simulations. The horizontal axis denotes the value of $p$,
and the vertical axis the initial strategy $\mathbf{x}=(x,1-x)$ in
the population. At each time step, a randomly generated mutant
strategy $\mathbf{v}=(v,1-v)$ for $v \in [0,1]$ will randomly appear
in the population with probability $0.01$. The color of each point
on the $p$ - $x$ plane represents the average proportion of the
initial strategy in the population after $10^4$ time steps in $100$
runs. The black dashed curve, the white dashed curve and the
boundary $x=1$ represent the theoretical predictions for three
strong SNE strategies as functions of $p$.}
    \label{Fig1}
\end{figure}

\emph{\textbf{Conclusion.}} Stochastic fluctuations (or uncertainty) in returns in a temporally varying environment could have a profound impact on the evolution of animal behavior. Therefore, introducing the concept of a stochastic Nash equilibrium (SNE) that extends the classical concept of a NE \cite{smi1982, hof1998} to take into account random payoffs and revealing its relationship with a stochastically evolutionarily stable (SES) strategy \cite{zhe2017, fen2022} may be of prime interest.

For the definition of a SNE as the strategy that is the best reply
to itself in a stochastic framework, we have to compare geometric
rather than arithmetic mean payoffs of strategies. Moreover, the SNE
is said weak in the case of an equality for all other strategies,
while it is said strong if there is a strict inequality for all
other strategies.

Considering a linear evolutionary game with a random payoff matrix $\mathbf{A}(t)$ at each time step $t\ge 1$ and using conditions for stochastic stability or instability of equilibria \cite{zhe2017}, we have shown that:
 (\textbf{\emph{i}}) at least one SNE exists; (\textbf{\emph{ii}}) a SES strategy must be a SNE; (\textbf{\emph{iii}}) a strong SNE must be a SES strategy, but this is not necessarily the case for a weak SNE; and (\textbf{\emph{iv}}) a strong SNE can be a completely mixed strategy, and more than one can exist.

The concept of a SNE defined in a stochastic framework not only fully covers the classical concept of a NE in a deterministic setting, but a SNE may have some properties that a NE cannot possess. For instance, in classical matrix games, a completely mixed strategy cannot be a strict NE (strong NE in our terminology), while a completely mixed NE must correspond to an interior equilibrium in the evolutionary dynamics of pure strategies \cite{smi1982, hof1998, bro2022}. On the contrary, a completely mixed strategy can be a strong SNE as shown in this paper, but it must not correspond to an interior constant equilibrium in the stochastic evolutionary dynamics of pure strategies \cite{zhe2017, fen2021, fen2022}.

The concept of a SNE, especially the existence of a completely mixed strong SNE that is noise-induced, may play an important role for a better understanding of the evolutionary complexity of animal behavior in natural populations subject to environmental noise, such as the evolution of cooperation in a stochastic environment \cite{fen2023, ber2006, per2006}. This is also consistent with Maynard Smith's \cite{smi1982} emphasis on the importance of mixed strategies in evolutionary games.

\renewcommand{\theequation}{A\arabic{equation}}
\setcounter{equation}{0}

\renewcommand{\thefigure}{A\arabic{figure}}
\setcounter{figure}{0}

\section*{Appendix}

Consider a population in which only two mixed strategies are in use, $\mathbf{u}=(u, 1-u)$ and $\mathbf{v}=(v, 1-v)$ with $u, v \in [0,1]$. The payoff matrix for these two mixed strategies at time step $t \ge 1$ is given by
\begin{align}\label{pmA}
\begin{pmatrix} \mathbf{u} \cdot \mathbf{A}(t) \mathbf{u} & \mathbf{u} \cdot \mathbf{A}(t) \mathbf{v} \\ \mathbf{v} \cdot \mathbf{A}(t) \mathbf{u} & \mathbf{v} \cdot \mathbf{A}(t) \mathbf{v} \end{pmatrix} \ .
\end{align}
This appendix provides a detailed analysis for the existence of
stochastic Nash equilibria as defined in Eq. (\ref{SNEbis}) in the
main text.

For convenience, we define
\begin{align}
Q(u,v) &= \left < \log \big (\mathbf{u} \cdot \mathbf{A}(t) \mathbf{v} \big) \right > \ .
\end{align}
Note that the expression $\mathbf{u} \cdot \mathbf{A}(t) \mathbf{v}$
is a convex combination of the elements of the payoff matrix
$\mathbf{A}(t)$ whose coefficients are $uv$, $u(1-v)$, $(1-u)v$ and
$(1-u)(1-v)$, respectively, and that the entries of $\mathbf{A}(t)$
are positive random variables that are uniformly bounded below and
above by some positive constants, that is, there exist real numbers
$A$ and $B$ such that $0< A \le a_{ij}(t) \le B$ for all $i,j=1,2$
and $t \ge 1$. Thus, the function $\log \big (\mathbf{u} \cdot
\mathbf{A}(t) \mathbf{v} \big)$ is continuous and differentiable
with respect to $u$ and $v$. Moreover, the partial derivative of
$Q(u,v)$ with respect to $u$ is given by
\begin{align}\label{firstderivative}
\frac{\partial Q(u,v)}{\partial u} &=
\frac{\partial \left < \log \big (\mathbf{u} \cdot \mathbf{A}(t) \mathbf{v} \big) \right >}{\partial u} \nonumber \\
&= \left < \frac{\partial \log \big (\mathbf{u} \cdot \mathbf{A}(t) \mathbf{v} \big)}{\partial u} \right > \nonumber \\
&= \left < \frac{\big( \mathbf{A}(t) \mathbf{v} \big )_1 - \big( \mathbf{A}(t) \mathbf{v} \big )_2}{\mathbf{u} \cdot \mathbf{A}(t) \mathbf{v}} \right > \ .
\end{align}
Similarly, the second-order partial derivative of $Q(u,v)$ with
respect to $u$ exists and is given by
\begin{align}\label{second}
\frac{\partial^2 Q(u,v)}{\partial u^2} &= - \left < \frac{\big [ \big( \mathbf{A}(t) \mathbf{v} \big )_1 - \big( \mathbf{A}(t) \mathbf{v} \big )_2 \big ]^2}{\big (\mathbf{u} \cdot \mathbf{A}(t) \mathbf{v} \big )^2} \right > \ .
\end{align}
Note also that $0 < A^2\leq \big (\mathbf{u} \cdot \mathbf{A}(t) \mathbf{v} \big )^2\leq B^2$ and $\big [ \big( \mathbf{A}(t) \mathbf{v} \big )_1 - \big( \mathbf{A}(t) \mathbf{v} \big )_2 \big ]^2 \ge 0$. Thus, we have
\begin{align}\label{bound}
\frac{\partial^2 Q(u,v)}{\partial u^2} \leq - \frac{1}{B^2} \left < \big [ \big( \mathbf{A}(t) \mathbf{v} \big )_1 - \big( \mathbf{A}(t) \mathbf{v} \big )_2 \big ]^2 \right > \leq 0 \ .
\end{align}
 We can conclude that
$\partial^2 Q(u,v)/ \partial u^2=0$
 if and only if $\mathbf{P} \left ( \big( \mathbf{A}(t)
\mathbf{v} \big )_1 = \big( \mathbf{A}(t) \mathbf{v} \big )_2
\right)=1$, in which case we have also $\partial Q(u,v)/\partial u=0$.

Two cases have to be considered.

{\bf Case 1.} $v \in [0,1]$ is such that $\mathbf{P} \left ( \big(
\mathbf{A}(t) \mathbf{v} \big )_1 = \big( \mathbf{A}(t) \mathbf{v}
\big )_2 \right)=1$, so that $\mathbf{u} \cdot \mathbf{A}(t)
\mathbf{v}=\mathbf{v} \cdot \mathbf{A}(t) \mathbf{v}$ with
probability $1$, from which
\begin{align}
Q(u,v) - Q(v,v) &= \left < \log \Big(\frac{\mathbf{u} \cdot \mathbf{A}(t) \mathbf{v} }{\mathbf{v}
\cdot \mathbf{A}(t) \mathbf{v} } \Big)\right > = 0
\end{align}
for all possible $\mathbf{u}=(u, 1-u)$ with $u \in [0,1]$. This
implies that $\mathbf{v}=(v, 1-v)$ is a weak SNE, but not a strong
SNE as defined in the main text. Moreover, note that the above
condition takes the form
\begin{align}\label{eq_v}
\mathbf{P} \left (\big( a_{11}(t)-a_{21}(t) \big)v = \big( a_{22}(t)-a_{12}(t) \big)(1-v) \right)=1.
\end{align}
A solution $v \in [0,1]$ that satisfies this condition involves four possible situations.

({\bf \emph{i}}) $\mathbf{P} \big (a_{11}(t)=a_{21}(t) \big )=1$ and $\mathbf{P} \left (a_{22}(t)=a_{12}(t) \right)=1$, in which case every $v \in [0,1]$ is a solution.

({\bf \emph{ii}}) $\mathbf{P} \big (a_{11}(t)=a_{21}(t) \big )=1$ and $\mathbf{P} \left
(a_{22}(t)=a_{12}(t) \right)<1$, in which case $v=1$ is the unique solution.

({\bf \emph{iii}}) $\mathbf{P} \left (a_{11}(t)=a_{21}(t) \right)<1$ and $\mathbf{P} \big (a_{22}(t)=a_{12}(t) \big )=1$,
in which case $v=0$ is the unique solution.

({\bf \emph{iv}}) $\mathbf{P} \left (a_{11}(t)=a_{21}(t) \right)<1$ and $\mathbf{P} \big (a_{22}(t)=a_{12}(t) \big )<1$, in which case $r>0$ satisfying $\mathbf{P} \big (
a_{11}(t)-a_{21}(t)=r\left(a_{22}(t)-a_{12}(t)\right) \big )=1$ is unique if it exists, and then $v=1/(1+r) \in (0, 1)$ is the unique solution.

As for stochastic local stability, it is known that, under the condition $\mathbf{u} \cdot \mathbf{A}(t) \mathbf{v}=\mathbf{v} \cdot
\mathbf{A}(t) \mathbf{v}$ with probability $1$ in the payoff matrix (\ref{pmA}), the mixed strategy $\mathbf{v}$ is SLS against the mixed strategy $\mathbf{u}$ if
\begin{align}\label{SNE_tobeSES}
\left < \frac{\mathbf{v} \cdot \mathbf{A}(t) \mathbf{u}}{\mathbf{v} \cdot \mathbf{A}(t) \mathbf{v}} -\frac{\mathbf{u} \cdot \mathbf{A}(t) \mathbf{u}}{\mathbf{v} \cdot
\mathbf{A}(t) \mathbf{v}} \right > = \left < \frac{\mathbf{v} \cdot \mathbf{A}(t) \mathbf{u}}{\mathbf{v} \cdot \mathbf{A}(t) \mathbf{v}} \right > - \left < \frac{\mathbf{u} \cdot \mathbf{A}(t) \mathbf{u}}{\mathbf{v} \cdot \mathbf{A}(t) \mathbf{v}} \right > > 0
\end{align}
and SLU if the inequality is reversed (see Eq. (15) in
\cite{zhe2017}). Besides, $\mathbf{v}$ is stochastically
evolutionarily stable (SES) if it is SLS against all $\mathbf{u} \ne
\mathbf{v}$.

Note that, using Eq. (\ref{eq_v}) and introducing the notation
$\mathbf{z}=(1, -1)$, we have almost surely
\begin{align}
(\mathbf{v}-\mathbf{u}) \cdot \mathbf{A}(t) \mathbf{u} &= (v-u) \mathbf{z} \cdot \mathbf{A}(t) \mathbf{u} \nonumber \\
&= (v-u) \Big [ u \big (a_{11}(t) -a_{21}(t) \big) +(1-u) \big (a_{12}(t) -a_{22}(t) \big ) \Big ] \nonumber \\
&= (v-u) \Big [ u \big (a_{11}(t) -a_{21}(t) \big) -v \big (a_{11}(t) -a_{21}(t) \big) \nonumber \\
& \ \ \ \ \ \ \ +(1-u) \big (a_{12}(t) -a_{22}(t) \big ) +(1-v) \big (a_{22}(t) -a_{12}(t) \big ) \Big ] \nonumber \\
&= -(v-u)^2 \mathbf{z} \cdot \mathbf{A}(t) \mathbf{z}
\end{align}
 and
\begin{align}\label{vAv}
&\big ( \mathbf{z} \cdot \mathbf{A}(t) \mathbf{z}\big )^2 \mathbf{v} \cdot \mathbf{A}(t) \mathbf{v} \nonumber\\
&= \big ( a_{11}(t) -a_{12}(t) -a_{21}(t) +a_{22}(t) \big )^2 \nonumber\\
&\quad \times \left[ v^2a_{11}(t) + v(1-v)\left(a_{12}(t)+a_{21}(t)\right) + (1-v)^2a_{22}(t)\right]\nonumber\\
&= \left(a_{22}(t)-a_{12}(t)\right)^2a_{11}(t)
 +\left(a_{22}(t)-a_{12}(t)\right)\left(a_{11}(t)-a_{21}(t)\right)\left(a_{12}(t)+a_{21}(t)\right)\nonumber\\
 &\quad +\left(a_{11}(t)-a_{21}(t)\right)^2a_{22}(t)\nonumber\\
 &=\big ( \mathbf{z} \cdot \mathbf{A}(t) \mathbf{z} \big ) \det \mathbf{A}(t) \ ,
\end{align}
where
\begin{align}\label{zAz}
\mathbf{z} \cdot \mathbf{A}(t) \mathbf{z}=a_{11}(t) -a_{12}(t) -a_{21}(t) +a_{22}(t)
\end{align}
and
$\det \mathbf{A}(t)=a_{11}(t)a_{22}(t)-a_{12}(t)a_{21}(t)$. Moreover, $\det \mathbf{A}(t)=0$ almost surely if $\mathbf{z} \cdot \mathbf{A}(t) \mathbf{z}=0$, since then $a_{22}(t) -a_{12}(t)=a_{11}(t) -a_{21}(t)=0$ almost surely. Therefore, Eq. (\ref{vAv}) can be replaced by
\begin{align}\label{vAvbis}
\big ( \mathbf{z} \cdot \mathbf{A}(t) \mathbf{z}\big ) \mathbf{v} \cdot \mathbf{A}(t) \mathbf{v}
 = \det \mathbf{A}(t)
\end{align}
with $\det \mathbf{A}(t)=0$ if and only if $\mathbf{z} \cdot \mathbf{A}(t) \mathbf{z}=0$. We can conclude that
\begin{align}
\left < \frac{\mathbf{v} \cdot \mathbf{A}(t) \mathbf{u}}{\mathbf{v} \cdot \mathbf{A}(t) \mathbf{v}} -\frac{\mathbf{u} \cdot \mathbf{A}(t) \mathbf{u}}{\mathbf{v} \cdot
\mathbf{A}(t) \mathbf{v}} \right >=\left <\frac{(\mathbf{v}-\mathbf{u}) \cdot \mathbf{A}(t) \mathbf{u}}{\mathbf{v} \cdot \mathbf{A}(t) \mathbf{v}} \right > &= -(v-u)^2 D \ ,
\end{align}
where
\begin{align}\label{D}
D &= \left < \frac{\big ( \mathbf{z} \cdot \mathbf{A}(t)
\mathbf{z}\big )^2}{\det \mathbf{A}(t)} \right >
\end{align}
with the convention that $ \big (\mathbf{z} \cdot \mathbf{A}(t) \mathbf{z}\big )^2/\det \mathbf{A}(t)=0$ when $\mathbf{z} \cdot \mathbf{A}(t) \mathbf{z}=\det \mathbf{A}(t)=0$.
If $D<0$, then the inequality in Eq. (\ref{SNE_tobeSES}) holds for all $\mathbf{u} \ne \mathbf{v}$, which means that $\mathbf{v}$ is SES. On the contrary, this is not possible when $D>0$,

{\bf Case 2.} $v \in [0,1]$ is such that $\mathbf{P} \left ( \big(
\mathbf{A}(t) \mathbf{v} \big )_1 = \big( \mathbf{A}(t) \mathbf{v}
\big )_2 \right)<1$, from which
\begin{align}
\left < \big [ \big( \mathbf{A}(t) \mathbf{v} \big )_1 - \big( \mathbf{A}(t) \mathbf{v} \big )_2 \big ]^2 \right > > 0,
\end{align}
and then Eq. (\ref{second}) yields $\partial^2 Q(u,v)/\partial u^2 <
0$. This means that $Q(u,v)$ is a strictly concave function with
respect to $u \in [0,1]$, which then reaches a unique global maximum
at some point
\begin{align}
\phi (v) = \mathop{\arg\max}\limits_{u \in [0,1]} Q(u,v)\in [0, 1] \ .
\end{align}
On the other hand, it is known that the mixed strategy
$\mathbf{v}$ is SLS
if
$Q(u,v) < Q(v,v)$, and SLU if the inequality is reversed (see Eq. (10) in \cite{zhe2017}). Therefore, we have
\begin{align}
Q(u,v) - Q(v,v) &= \left < \log \Big(\frac{\mathbf{u} \cdot \mathbf{A}(t) \mathbf{v}}{\mathbf{v} \cdot \mathbf{A}(t) \mathbf{v} } \Big) \right > \le 0
\end{align}
with an equality to $0$ if and only if $\mathbf{u} = \mathbf{v}$,
and then $\mathbf{v}$ is SES, only when $v=\phi(v)$. In this case,
we have
\begin{align}
\frac{\partial Q(u,v)}{\partial u} \Big |_{u=v}
\begin{cases}
\le 0& \textrm{ if } v=0,\\
\ge 0& \textrm{ if } v=1,\\
=0 &\textrm{ if } 0 < v<1,
\end{cases}
\end{align}
and $\mathbf{v}=(v, 1-v)$ is a strong SNE as defined in the main
text.

As for the existence of a strong SNE, two situations have to be considered.

({\bf \emph{i}}) $\mathbf{P} \left ( \big( \mathbf{A}(t) \mathbf{v}
\big )_1 = \big( \mathbf{A}(t) \mathbf{v} \big )_2 \right)<1$ for
all $v \in [0,1]$, so that $\phi(v)$ is well defined and continuous
on $[0, 1]$. Then, this is also the case for the function defined by
$g(v)=\phi(v)-v$. Moreover, we have $g(0)=\phi(0) \ge 0$ and
$g(1)=\phi(1)-1 \le 0$. According to the mean value theorem, there
exists $v^* \in [0,1]$ such that $g(v^*)=0$, that is,
$\phi(v^*)=v^*$. The corresponding mixed strategy
$\mathbf{v^*}=(v^*, 1-v^*)$ is then a strong SNE.

({\bf \emph{ii}}) $\mathbf{P} \left ( \big( \mathbf{A}(t)
\mathbf{\hat{v}} \big )_1 = \big( \mathbf{A}(t) \mathbf{\hat{v}}
\big )_2 \right)=1$ for some $\hat{v} \in [0,1]$, so that
$\mathbf{\hat{v}}=(\hat{v}, 1-\hat{v})$ is a weak SNE since
$Q(u,\hat{v})-Q(\hat{v},\hat{v})=0$ for all $u \in [0,1]$, in which
case $\phi(\hat{v})= [0,1]$. From the analysis in Case 1, we know
that $\hat{v}$ is unique if it exists unless $\mathbf{P} \big
(a_{11}(t)=a_{21}(t) \big )=1$ and $\mathbf{P} \left
(a_{22}(t)=a_{12}(t) \right)=1$, which is excluded here since we
assume that there exists $v \in [0,1]$ such that $\mathbf{P} \left (
\big( \mathbf{A}(t) \mathbf{v} \big )_1 = \big( \mathbf{A}(t)
\mathbf{v} \big )_2 \right)<1$. Note also that, if $\hat{v}$ exists,
then $\phi(v)$ is well defined and continuous on the intervals $[0
,\hat{v})$ and $(\hat{v},1]$.

From Eqs. (\ref{firstderivative}) and (\ref{zAz}), we find that $F(u,v)=\partial Q(u,v)
\big / \partial u$ has a partial derivative with respect to $v$ given by
\begin{align}\label{dFdv}
\left < \frac{ (\mathbf{z} \cdot \mathbf{A}(t)
\mathbf{z})(\mathbf{u} \cdot \mathbf{A}(t) \mathbf{v}) - \left(\big(
\mathbf{A}(t) \mathbf{v} \big )_1 - \big( \mathbf{A}(t) \mathbf{v}
\big )_2\right) \left(\big(\mathbf{u}^T \mathbf{A}(t) \big )_1 -
\big( \mathbf{u}^T\mathbf{A}(t) \big )_2\right) }{\big (\mathbf{u}
\cdot \mathbf{A}(t) \mathbf{v} \big )^2} \right > .
\end{align}
Evaluating at $v=\hat{v}$ and using Eq. (\ref{vAvbis}) for the expression of $\mathbf{u} \cdot \mathbf{A}(t) \mathbf{\hat{v}}=\mathbf{\hat{v}} \cdot \mathbf{A}(t) \mathbf{\hat{v}}$, we get
\begin{align}
\frac{\partial F(u,\hat{v}) }{\partial v} = D
\end{align}
with $D$ as defined in Eq. (\ref{D}), while we have $F(u, \hat{v})=0$ owing to Eq. (\ref{firstderivative}), for all $u\in [0, 1]$.

If $D>0$ and $\hat{v} >0$, then $\partial F(1,\hat{v}) /\partial v >0$ and there exists $\epsilon_1>0$ such that
$F(1,v) < 0$ for $v \in (\hat{v}-\epsilon_1, \hat{v})$. On the other hand, from Eq. (\ref{bound}) and the unicity of $\hat{v}$, the function $F(u,v)$ is strictly increasing
with respect to $u$ for $v \ne \hat{v}$, which implies that $F(u,v) \le F(1,v)<0$ for $u\in [0, 1]$ and $v \in (\hat{v}-\epsilon_1, \hat{v})$. This also implies that $Q(u,v)$ is a strictly decreasing function of $u$ whose maximum is reached at $u=0$ for $v \in (\hat{v}-\epsilon_1, \hat{v})$, in which case $\phi(v)=0$. Therefore, we have $g(\hat{v}^-)=\lim \limits_{v \uparrow \hat{v}} (\phi(v)-v)=-\hat{v}<0$, while $g(0)=\phi(0) \ge 0$. The mean value theorem applied to the continuous function $g(v)$ on $[0, \hat{v})$ ensures the existence of $v^*_1 \in [0, \hat{v})$ such that $g(v^*_1)=0$, that is, $\phi(v^*_1)=v^*_1$. The corresponding mixed strategy $\mathbf{v}^*_1=(v^*_1, 1-v^*_1)$ is then a strong SNE.

Analogously, if $D>0$ and $\hat{v} <1$, then we can find $\epsilon_2>0$ such that $F(u,v) \ge F(1,v)>0$ for $u\in [0, 1]$ and $v \in (\hat{v}, \hat{v}+\epsilon_2)$, which implies $\phi(v)=1$. In this case, we have $g(\hat{v}^+)=\lim \limits_{v \downarrow \hat{v}} (\phi(v)-v) =1-\hat{v}>0$, while $g(1)=\phi(1)-1 \le 0$. Therefore, there exists $v^*_2 \in (\hat{v}, 1]$ such that $g(v^*_2)=0$, that is, $\phi(v^*_2)=v^*_2$, and the corresponding mixed strategy $\mathbf{v}^*_2=(v^*_2, 1-v^*_2)$ is a strong SNE.

Note that, if $D>0$ and $\hat{v}\in (0,1)$, then
there are at least three stochastic Nash equilibria, which are $\hat{\mathbf{v}}=(\hat{v},
1-\hat{v})$, $\mathbf{v}^*_1=(v^*_1, 1-v^*_1)$ and
$\mathbf{v}^*_2=(v^*_2, 1-v^*_2)$, respectively, with
$v^*_1<\hat{v}<v^*_2$. Whereas $\hat{\mathbf{v}}$ is a SNE that is not
SES (see Case 1), the mixed strategies $\mathbf{v}^*_1$ and $\mathbf{v}^*_2$ are both strong
SNE and then necessarily SES.

Similarly, if $D>0$ and $\hat{v}=0$ or $1$, then there is at least
one SNE apart from $\hat{v}$, which is a strong SNE, and then SES,
contrary to $\hat{v}$.

If $D<0$, however, it is possible that there is no other SNE except
for $\hat{\mathbf{v}}$, which is a weak SNE that is SES.

Finally, in the case where $D=0$, for which $\partial F(u,\hat{v}) /\partial v=0$ for all
$u \in [0,1]$,
we consider the second-order partial derivative of $F(u,v)$ with respect to $v$ evaluated at $v=\hat{v}$, which is given by
\begin{align}
\frac{\partial^2 F(u,\hat{v}) }{\partial v^2} &=
-2\left < \frac{ (\mathbf{z} \cdot \mathbf{A}(t) \mathbf{z}) \left(\big(\mathbf{u}^T \mathbf{A}(t) \big )_1 - \big( \mathbf{u}^T\mathbf{A}(t) \big )_2\right) }{\big (\mathbf{u} \cdot \mathbf{A}(t) \mathbf{\hat{v}} \big )^2} \right > \nonumber\\
&= -2\left < \frac{ (\mathbf{z} \cdot \mathbf{A}(t) \mathbf{z})^3
\left(\big(\mathbf{u}^T \mathbf{A}(t) \big )_1 - \big(
\mathbf{u}^T\mathbf{A}(t) \big )_2\right) }{\big ( \det
\mathbf{A}(t) \big )^2} \right >
\end{align}
owing to Eqs. (\ref{dFdv}) and (\ref{vAvbis}) with the convention that $ \big (\mathbf{z} \cdot \mathbf{A}(t) \mathbf{z}\big )^3/(\det \mathbf{A}(t))^2=0$ when $\mathbf{z} \cdot \mathbf{A}(t) \mathbf{z}=\det \mathbf{A}(t)=0$. Defining
\begin{subequations}
\begin{align}
G&= \left < \frac{(\mathbf{z} \cdot \mathbf{A}(t) \mathbf{z})
^3}{\big ( \det \mathbf{A}(t) \big )^2} \big (a_{11}(t) -a_{12}(t)
\big ) \right >
 \ , \\
H&= \left < \frac{(\mathbf{z} \cdot \mathbf{A}(t) \mathbf{z})
^3}{\big ( \det \mathbf{A}(t) \big )^2} \big (a_{21}(t) -a_{22}(t)
\big ) \right > \ ,
\end{align}
\end{subequations}
we find that
\begin{align}
G-H &= \left < \frac{(\mathbf{z} \cdot \mathbf{A}(t) \mathbf{z})
^4}{\big ( \det \mathbf{A}(t) \big )^2} \right > > 0.
\end{align}
Here, we use the fact $\mathbf{P} \left ( \mathbf{z} \cdot
\mathbf{A}(t) \mathbf{z}=0 \right)<1$, since it is assumed that
there exists $v \in [0,1]$ such that $\mathbf{P} \left ( \big(
\mathbf{A}(t) \mathbf{v} \big )_1 = \big( \mathbf{A}(t) \mathbf{v}
\big )_2 \right)<1$. Moreover, we have
\begin{align}
\frac{d^2 F(u, \hat{v})}{dv^2} =-2 \big [ uG +(1-u) H \big ]
\end{align}
for $u \in [0,1]$.

There are three cases to consider:

\begin{enumerate}
\item if $0\le H<G$, then we have $d^2 F(u, \hat{v})/dv^2<0$ for $u \in (0,1]$, from which $\phi(\hat{v}^-)=\lim \limits_{v \uparrow \hat{v}} \phi(v)=0$ if $\hat{v}>0$ and $\phi(\hat{v}^+)=\lim \limits_{v \downarrow \hat{v}} \phi(v)=0$ if $\hat{v}<1$, since then $F(u,v)\le F(1,v) <0$ for $u \in (0,1]$ and $\phi(v)=0$ for $v\ne \hat{v}$ close enough to $\hat{v}$;
\item if $H<G\le 0$, then we have $d^2 F(u,\hat{v})/dv^2>0$ for $u \in [0,1)$, from which $\phi(\hat{v}^-)=1$ if $\hat{v}>0$ and $\phi(\hat{v}^+)=1$ if $\hat{v}<1$ by symmetry with the previous case with $F(u,v)\ge F(0,v) >0$ for $u \in [0,1)$ and $\phi(v)=1$ for $v\ne \hat{v}$ close enough to $\hat{v}$; and
\item if $G > 0 > H$, then we have $d^2 F(\hat{u}, \hat{v})/dv^2=0$ for $\hat{u}=H \big /(H-G) \in (0,1)$, while $d^2 F(u, \hat{v})/dv^2<0$ for $u\in (\hat{u},1]$ and $d^2 F(u, \hat{v})/dv^2>0$ for $u\in [0, \hat{u})$, from which $\phi(\hat{v}^-)=\phi(\hat{v}^+)=\hat{u}$ by analogy with the two previous cases.
\end{enumerate}
Now, let us define $g(v)= \phi(v)-v$ for $v\ne \hat{v}$ and
\begin{align}
g(\hat{v})=
\begin{cases}
-\hat{v} &\textrm{ if } \hat{u}\le 0, \\
1 - \hat{v} & \textrm{ if }\hat{u}\ge 1,\\
\hat{u}- \hat{v} & \textrm{ if }\hat{u} \in (0,1).
\end{cases}
\end{align}
This is a continuous function on $ [0, 1]$ in the three cases above
with $g(0)\ge 0$ and $g(1)\le 0$. Moreover, applying the mean value
theorem, there exist $v^*_1 \in [0, \hat{v})$ such that $g(v^*_1)=0$
if $\hat{v}> \max (0, \hat{u})$, as well as $v^*_2 \in (\hat{v},1]$
such that $g(v^*_2)=0$ if $\hat{v}< \min (1, \hat{u})$. The
corresponding mixed strategies $\mathbf{v}^*_1=(v^*_1, 1-v^*_1)$ and
$\mathbf{v}^*_2=(v^*_2, 1-v^*_2)$ are strong SNE, while
$\mathbf{\hat{v}}=(\hat{v}, 1-\hat{v})$ is a SNE that is not SES.

We now give an example to show the nature of the function $\phi(v)$
according to the sign of $D$.

\textbf{Example A1.} Consider a random payoff matrix
\begin{align}\label{exS1}
\mathbf{A}(t)=\begin{pmatrix} c+r \eta_t & b \\ c & b+\eta_t
\end{pmatrix} \ ,
\end{align}
where $b,c,r$ are positive constants, and
\begin{align}
\eta_t=\begin{cases}
s &\textrm{ with probability } p \ , \\
-s & \textrm{ with probability } 1-p \ ,
\end{cases}
\end{align}
with $0<s< \min (b,c/r)$. It is easy to show that
$(\hat{v},1-\hat{v})$ with $\hat{v}=1/(1+r) \in (0,1)$ is a unique
weak SNE. Moreover, from Eq. (\ref{D}), we have
\begin{align}
D&= \left < \frac{(r+1)^2 \eta_t}{c+rb+r \eta_t} \right
>=s\frac{(2p-1)(c+rb)-rs}{(c+rb)^2-r^2s^2} \ .
\end{align}
Since $c+rb>rs$, we can see that the sign of $D$ corresponds to the
sign of $(2p-1)(c/r+b)-s$. Moreover,
\begin{align}
F(u,v)&=\frac{\partial Q(u,v)}{\partial u}=\frac{v-\hat{v}}{\hat{v}}
\left < \frac{\eta_t}{\mathbf{u} \cdot \mathbf{A}(t) \mathbf{v}}
\right
> \nonumber \\
&=\frac{v-\hat{v}}{\hat{v}} \left <
\frac{\eta_t}{(vc+(1-v)b)+(uvr+(1-u)(1-v))\eta_t} \right
> \ .
\end{align}
By solving the equation $F(u,v)=0$ for $v \neq \hat{v}$, which can
be simplified to a one-dimensional equation, we get
\begin{align}
\phi(v)&=\begin{cases}
\min \left (1,\max \left (0,\frac{((2p-1)(c-b)+s)v+(2p-1)b-s}{s(r+1)v-s} \right ) \right ) &\textrm{ for } v \neq \hat{v} \ , \\
[0,1] & \textrm{ for } v=\hat{v} \ .
\end{cases}
\end{align}
If $D>0$, then $\phi(v)$ has at least two intersection points with
the line $\phi(v)=v$ that correspond to two strong SNE (see Fig.
A1a). If $D<0$, then a strong SNE may exist or not (see Fig. A1b and
A1c). On the other hand, if $D=0$, which means that
$s=(2p-1)(c/r+b)$, then $\phi(v)$ becomes the constant
$c/(c+rb)=\hat{u} \in (0,1)$ for $v \neq \hat{v}$. If $\hat{u} \neq
\hat{v}$, that is, $b \neq c$, then we can always find a strong SNE
given by $(\hat{u},1-\hat{u})$. In this case, the mixed strategy
$(\hat{u},1-\hat{u})$ is the unique best reply to any other
strategy, except for $(\hat{v},1-\hat{v})$, with respect to the
geometric means of the payoffs (see Fig. A1d). If $b=c$, then
$(\hat{v},1-\hat{v})$ is the unique SNE in the system.

\begin{figure}
    \begin{center}
    \includegraphics[height=7.2cm,width=8.47cm]{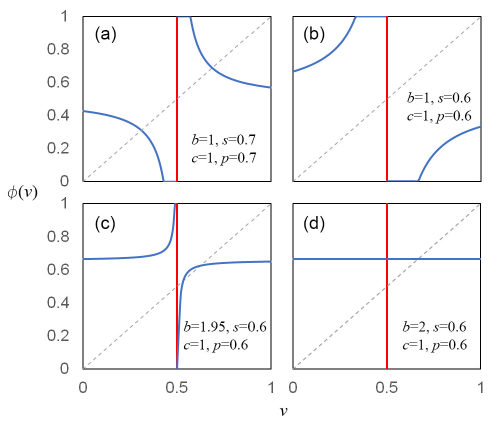}
    \end{center}
    \caption{\textbf{The function $\phi(v)$ in Example A1}}
    \label{FigA1}
\end{figure}

\textbf{Example A2.} In this example, we consider a random payoff
matrix
\begin{align}\label{exS1}
\mathbf{A}(t)=\begin{pmatrix} c+(1-a)(1-\hat{v})\eta_t & b-(1-a)\hat{v}\eta_t \\
c-a(1-\hat{v})\eta_t & b+a\hat{v}\eta_t
\end{pmatrix} \ ,
\end{align}
where $b,c$ are positive constants and $a,\hat{v} \in (0,1)$, while
$\eta_t$ is any non-constant white noise with
$\left<\eta_t\right>=0$ that makes the entries in $\mathbf{A}(t)$
always positive (for instance, $|\eta_t|<\delta$, where $\delta>0$
is small enough). It is still easy to see that $(\hat{v},1-\hat{v})$
is the unique weak SNE. From Eq.(\ref{D}), we get
\begin{align}
D&= \left < \frac{(1-2\hat{v})^2}{(1-\hat{v})b+\hat{v}c}\eta_t
\right
>=\frac{(1-2\hat{v})^2}{(1-\hat{v})b+\hat{v}c} \left < \eta_t \right
> =0
\end{align}
almost surely, and
\begin{align}
F(u,v)&=(v-\hat{v}) \left < \frac{\eta_t}{\mathbf{u} \cdot
\mathbf{A}(t) \mathbf{v}} \right
> \nonumber \\
&=(v-\hat{v}) \left <
\frac{\eta_t}{(v-\hat{v})(u(1-2a)-a)\eta_t+c\hat{v}+b(1-\hat{v})}
\right
> \ .
\end{align}
When $a \in (0,1/3)$, we have $\hat{u}=a/(1-2a) \in (0,1)$ and
$F(\hat{u},v)=\frac{v-\hat{v}}{c\hat{v}+b(1-\hat{v})} \left <\eta_t
\right>=0$, from which $\phi(v)=\hat{u}$ for $v \neq \hat{v}$. This
corresponds to the case where $D=0$, $G>0>H$. When $a \in
[1/3,1/2)$, we have $\hat{u}=1$ and this corresponds to the case
where $D=0$, $0 \le H < G$. Finally, when $a \in (1/2,1)$, we have
$\hat{u}=0$, which corresponds to the case where $D=0$, $ H < G \le
0$.

\section*{Acknowledgements}

\textbf{Funding:} In this study, C.L. was supported by the National
Natural Science Foundation of China (Grant No. 32271553) and the
Fundamental Research Funds for the Central Universities; T-J.F.,
X-D.Z. and Y.T. were supported by the National Natural
Science Foundation of China (Grants No. 32071610 and No. 31971511); S.L. was supported by the Natural Sciences and Engineering Research Council of Canada (Grant No. 8833).\\


\bibliographystyle{unsrt}

\end{document}